\documentclass[aps,pra,onecolumn,showpacs,notitlepage,groupedaddress]{revtex4-1}  
\setcitestyle{square}
\usepackage{graphicx}  
\usepackage{float}
\usepackage{dcolumn}   
\usepackage{bm}        
\usepackage{amssymb}   
\usepackage{amsmath}
\usepackage{isomath}
\usepackage[usenames, dvipsnames]{color}
\usepackage{xcolor}
\usepackage{soul}
\usepackage{physics}
\usepackage[normalem]{ulem}

\begin{document}

\widetext
\title{Subwavelength and directional control of flexural waves \\ in zone-folding induced topological plates}

\author{Rajesh Chaunsali}
\affiliation{Aeronautics and Astronautics, University of Washington, Seattle, WA, USA, 98195-2400}
\author{Chun-Wei Chen}
\thanks{Equally contributed first author}
\affiliation{Aeronautics and Astronautics, University of Washington, Seattle, WA, USA, 98195-2400}
\author{Jinkyu Yang}
\email{jkyang@aa.washington.edu}
\affiliation{Aeronautics and Astronautics, University of Washington, Seattle, WA, USA, 98195-2400}
\date{\today}

\begin{abstract}
Inspired by the quantum spin Hall effect shown by topological insulators, we propose a plate structure that can be used to demonstrate the pseudo-spin Hall effect for flexural waves. The system consists of a thin plate with periodically arranged resonators mounted on its top surface. We extend a technique based on the plane wave expansion method to identify a double Dirac cone emerging due to the zone-folding in frequency band structures. This particular design allows us to move the double Dirac cone to a lower frequency than the resonating frequency of local resonators. We then manipulate the pattern of local resonators to open subwavelength Bragg band gaps that are topologically distinct. Building on this method, we verify numerically that a waveguide at an interface between two topologically distinct resonating plate structures can be used for guiding low-frequency, spin-dependent one-way flexural waves along a desired path with bends. 
\end{abstract}
\maketitle

\section{Introduction}
A topological insulator has emerged as a new state of matter in condensed matter physics. This is a special type of insulator that conducts electricity only on its boundary. Here topology is relevant because one can predict the boundary properties of these \textit{finite} materials (i.e., finite-sized lattices) solely by knowing the bulk properties of \textit{infinite} materials (i.e., infinitely large lattices). Topological framework provides an elegant way to categorize the bulk properties in terms of a topological invariant, and thus, one expects a \textit{topological protection} and a degree of \textit{robustness} for the boundary properties \citep{Hasan2010,Qi2011}.

It is recent that this whole framework dealing with the flow of electrons has evolved further and influenced other areas such as photonics  \citep{Lu2014} and acoustics \citep{Xiao2015, Yang2015, Lu2016, He2016, Mei2016, Fleury2016, Zhang2017, Yves2017}. It has also propelled new design paradigm for artificial mechanical structures, so-called topological mechanical metamaterials, to achieve unconventional static (zero frequency) \citep{Kane2013, Paulose2014, Rocklin2016, Stenull2016, Bilal2017} and dynamic (non-zero frequency) \citep{Huber2015, Nash2015, Wang2015, Mousavi2015, Kariyado2015, YWang2015, Pal2016a, Pal2016b, Vila2017, Brendel2017Valley, Brendel2017Spin, Yu2017, Salerno2016, Weinstein2014, Chaunsali2016, Chaunsali2017, Prodan2017} responses. In particular, tailoring non-zero frequency responses, i.e., elastic waves in structures, on topological grounds shows tremendous potential to be used for energy harvesting, sensing, and impact mitigation purposes \citep{Huber2016}.

One of the most unique topological effects is the quantum spin Hall effect, the underlying phenomenon shown by topological insulators \citep{Kane2005, Bernevig2006}. These systems are passive in the sense that they do not require any external field, but still possess \textit{directional} boundary states. This is due to the presence of Kramers partners, i.e., two opposite spins of electron, which travel in the opposite directions on their boundaries, thereby keeping the time reversal symmetry intact. Although mechanical counterparts, being bosonic systems, do not possess these intrinsic spins, one can carefully design the system to have two pseudo-spins by imposing certain symmetries in the lattice, and thus realize the pseudo-spin Hall effect \citep{Huber2015, Mousavi2015, Pal2016a, Brendel2017Spin, Yu2017}. 

While previous studies have successfully reported the feasibility of the pseudo-spin Hall effect in mechanical settings, in this study, we focus on the feasibility of the same in less-explored continuum structures such as plates. One of the approaches that has been recently applied in plate structures is a so-called zone-folding technique \citep{Wu2015}, in which one rather considers a larger unit cell than an irreducible one in a hexagonal lattice arrangement, so that the frequency band structure folds onto itself, creating a \textit{double} Dirac cone at the $\mathrm{\Gamma}$ point. Based on the same, Brendel \textit{et al}. \citep{Brendel2017Spin} and Yu \textit{et al}. \citep{Yu2017} showed that purely geometric manipulation of holes can invoke topological effects in plates. However, these topological effects have been restricted to high-frequency wave modes. Therefore, in this research, we ask the question: How can one invoke the pseudo-spin Hall effect at \textit{low frequencies} for a given plate dimension? It is important because of several reasons, including (1) the low-frequency plate modes, such as flexural modes, carry a large amount of energy, and manipulating them could lead to relevant engineering applications, and (2) these lower modes generally require bigger lattice patterns of holes on conventional plates due to the Bragg condition, and thus, an improved way of controlling the low-frequency wave modes can relax the current stringent size limitations. Therefore, it would be a significant advancement to the current research trend if one can demonstrate low-frequency pseudo-spin Hall effect in a continuum mechanical structure such as plates, which are ubiquitous in many engineering disciplines.      
 
To address the aforementioned challenges, we propose a topological plate system that consists of a thin plate with periodically arranged local resonators mounted on its top surface. This locally resonant (LR) plate is a reminiscence of sonic crystals \citep{Liu2000}. Pal \textit{et al}. \cite{Pal2016b} proposed such a structure for realizing the elastic analogue of quantum valley Hall effect. Building on the similar methodology, in this research, we employ a technique based on the combination of the classical plate theory and the plane wave expansion (PWE) method \citep{Sigalas1994, Yu2006, Xiao2012}, which enables fast and efficient calculation of the wave dispersion relation. Furthermore, we integrate into this scheme the zone-folding technique to create a double Dirac cone for flexural wave modes. As a result, we report that the double Dirac cone can be formed in low-frequency regimes by tuning the resonating frequency of the resonators. In this way, we can acquaint a subwavelength characteristic of the proposed plate system, i.e., the lattice size of the LR plate being smaller than the wavelengths in the bare plate at operating frequencies. We then show that a purely geometric manipulation of the local resonator pattern results in the opening of a topologically trivial and non-trivial subwavelength Bragg band gaps around the double Dirac cone. Building on these findings, we verify numerically---by using the finite element method (FEM)---that a waveguide created at the interface of topologically distinct LR plates can guide low-frequency flexural waves along a designed path. Moreover, it shows a unique spin-depenedent one-way propagation characteristic. Unlike the traditional plate-based waveguides studied in the past \citep{Hsiao2007, Pennec2008, Pennec2009, Wu2009, Oudich2010, Casadei2012, Torrent2013, XWang2015, Jiang2015, Baboly2016, Chen2017}, we show that this LR topological plate system has potential to guide one-way flexural waves along a path with multiple bends---generally challenging in topologically trivial waveguides.  

The structure of this manuscript is as follows: in section II, we describe the design of the topological plate. In section III, we present the PWE method to calculate dispersion relation. In section IV, we show the zone-folding of bands and create a double Dirac cone in a subwavelength regime. In section V, we show the formation of a band gap around the double Dirac cone by perturbing the pattern of resonators on the LR plate. This facilitates the system to transition from a topologically trivial state to a non-trivial state. In section VI, we employ the FEM to show the existence of two local modes, each designated by a pseudo-spin (clockwise or counterclockwise), at the interface of topologically trivial and non-trivial lattices. In section VII, we demonstrate the feasibility of guiding low-frequency flexural wave modes along a path with bends and having a spin-dependent one-way propagation characteristic. In section VIII, we conclude this manuscript.

\section{Description of the Locally Resonant Topological Plate}
\begin{figure}[t]
\centering
\includegraphics[width=5in]{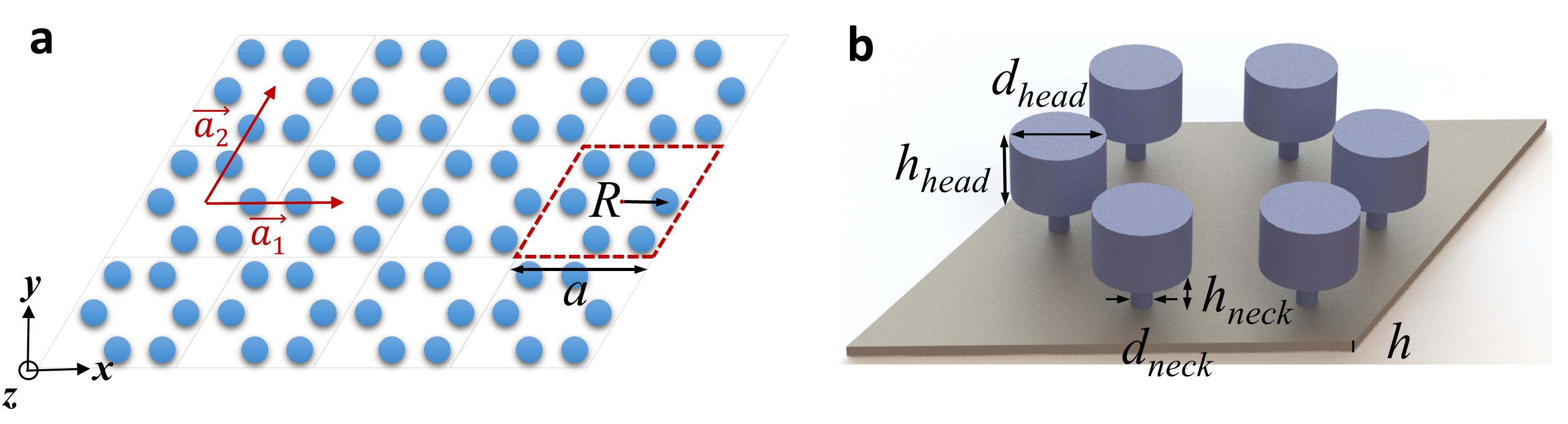}
\caption{(a) Lattice arrangement of the resonators with a rhombus shaped unit cell in red. Dimensional parameters and lattice vectors are also shown. (b) Detailed view of the unit cell with six resonators mounted on top of a thin plate.}
\label{fig1}
\end{figure}

Our system consists of a thin plate on which multiple local resonators are attached to form a lattice arrangement (Fig.~\ref{fig1}). The rhombus-shaped unit cell is of length $a$ and consists of six resonators in a hexagonal arrangement (Fig.~\ref{fig1}a). Each resonator is at a distance $R$ and rotationally symmetric from the center of the unit cell, showing the $C_6$ symmetry. $\vec{a}_1$ and $\vec{a}_2$ are the lattice vectors. We model the resonators as cylindrical heads attached to the plate with a thin neck (Fig.~\ref{fig1}b). In order to invoke the topological effects in the system, we will only vary radius $R$, keeping the $C_6$ symmetry intact in this unit cell. 

As a substrate material, we choose an aluminum plate ($E=77.6$ GPa, $\rho=2,730$ kg/m$^3$, $\nu=0.352$) of thickness $h=1$ mm with the unit cell of size $a=45$ mm. The resonator neck is made of acrylic plastic ($E_{neck}=3.2$ GPa) with $h_{neck}=5$ mm and $d_{neck}=2$ mm, whereas the resonator head is made of tungsten  ($\rho_{head}=19,260$ kg/m$^3$)   with $h_{head}=14$ mm and $d_{head}=9$ mm. The aforementioned material properties are based on nominal values of standard materials.

\section{Calculation of the unit cell dispersion}
We first evaluate dispersion characteristics of the unit cell design with six resonators for variable $R$. For fast calculations, we rely on the PWE method. To this end, we simplify the resonator design with a lumped mass ($m=\pi \rho_{head} d_{head}^2 h_{head}/4$) connected to the plate with a linear spring ($\beta=\pi E_{neck} d_{neck}^2/4h_{neck}$). In this process, we neglect the bending motion of the resonators and only consider their motion in the out-of-plane $z$-direction. The bending (and other modes) of resonators, though taken into account in full-scale models in later sections, do not affect the topological phenomenon in our system as those frequencies can be separated from the Dirac point we will be investigating and have minimal coupling with out-of-plane wave modes (see Appendix A). We have $h \ll a$, therefore the plate can be assumed to be thin \citep{Sigalas1994}, and the transverse motion of the plate can be calculated as per the classical plate theory (i.e., Kirchhoff-Love theory) \citep{Fahy2007}. Following the approaches taken by Pal \textit{et al.} \citep{Pal2016b}, Xiao \textit{et al.} \citep{Xiao2012}, and Torrent \textit{et al.} \citep{Torrent2013}, governing equations for the time-harmonic vibration of the unit cell with angular frequency $\omega$ can be written as 
\begin{subequations}
\begin{eqnarray}
D\nabla^4 w(\textbf{r}) -\omega^2 \rho h w(\textbf{r}) &=& -\beta \sum_{\alpha} [w(\textbf{R}_{\alpha})-\tilde{w}(\textbf{R}_{\alpha})]\delta(\textbf{r}-\textbf{R}_{\alpha}) \label{EOM1},\\
-\omega^2 m \tilde{w}(\textbf{R}_{\alpha}) &=& \beta [w(\textbf{R}_{\alpha})-\tilde{w}(\textbf{R}_{\alpha})], \label{EOM2}
\end{eqnarray}
\end{subequations}

\noindent where $D=Eh^3/12(1-\nu^2)$ represents the flexural rigidity of the plate, $\textbf{r}=(x,y) $ denotes the generalized coordinate of the plate, $w(\textbf{r})$ represents the transverse displacement of the plate, and $\tilde{w}(\textbf{R}_{\alpha})$ represents the displacement of the resonating masses attached at points $\textbf{R}_{\alpha}$. We have $\alpha=1,2,...,6$ for six different resonating masses per unit cell and $\delta(\textbf{r}-\textbf{R}_{\alpha})$ is a delta function in two dimensions. 

We introduce the following non-dimensional angular frequency
\begin{eqnarray}
 \mathrm{\Omega} = \omega  a^2 \sqrt{\frac{\rho h}{D}}. \nonumber
\end{eqnarray}

\noindent Also, the mass of the resonator can be normalized as  
\begin{eqnarray}
\gamma = \frac{m}{\rho A_c h}, \nonumber
\end{eqnarray}
\noindent where $A_c=\sqrt{3}~a^2/2$ is the area of a unit cell. We can write the normalized resonance frequency of a resonator as $\mathrm{\Omega}_r =a^2 \sqrt{(\beta/m) \rho h/D}$.

Employing the PWE method, we write the displacement of the plate for a Bloch wave vector $\textbf{K}$ as a superposition of multiple plane waves such that  
\begin{eqnarray} \label{PWE}
w(\textbf{r})=\sum_{\textbf{G}} W(\textbf{G})e^{-i(\textbf{K}+\textbf{G})\cdot\textbf{r}} \label{w},
\end{eqnarray}

\noindent where $W(\textbf{G})$ is a plane wave coefficient and $\textbf{G}$ denotes the reciprocal lattice vector given by $\textbf{G}=p\textbf{b}_1 +q\textbf{b}_2$, in which $p$ and $q$ are integers, and $\textbf{b}_1$ and $\textbf{b}_2$ are the basis vectors of the reciprocal lattice. We truncate the summation with respect to $\textbf{G}$ by choosing both $p$ and $q$ as $-M,-(M-1),...,0,...,(M-1), M$. Therefore, the reciprocal space is a $N\times N$ finite grid with $N=2M+1$. 

The displacement of the plate at the locations where the resonators are attached can be simply deduced from Eq.~\eqref{PWE} as
\begin{eqnarray}
w(\textbf{R}_{\alpha})&=&\sum_{\textbf{G}} W(\textbf{G})e^{-i(\textbf{K}+\textbf{G})\cdot\textbf{R}_{\alpha}}. \label{wa}
\end{eqnarray}

\noindent Substituting  Eq.~\eqref{w} and Eq.~\eqref{wa} into Eq.~\eqref{EOM1}  
\begin{eqnarray}
D \sum_{\textbf{G}'}\abs{\textbf{K}+\textbf{G}'}^4 W(\textbf{G}')e^{-i(\textbf{K}+\textbf{G}')\cdot\textbf{r}} &-&  \omega^2 \rho h \sum_{\textbf{G}'} W(\textbf{G}')e^{-i(\textbf{K}+\textbf{G}')\cdot\textbf{r}} \nonumber \\
&=& \beta \sum_{\alpha} \bigg[\tilde{w}(\textbf{R}_{\alpha})- \sum_{\textbf{G}'} W(\textbf{G}') e^{-i (\textbf{K}+\textbf{G}')\cdot\textbf{R}_{\alpha}} \bigg]\delta(\textbf{r}-\textbf{R}_{\alpha}). 
\end{eqnarray}

\noindent Multiplying both sides with $e^{i(\textbf{K}+\textbf{G})\cdot\textbf{r}}$, we obtain
\begin{eqnarray}
\sum_{\textbf{G}'} \bigg[D \abs{\textbf{K}+\textbf{G}'}^4-\omega^2 \rho h \bigg] W(\textbf{G}') e^{-i(\textbf{G}'-\textbf{G})\cdot\textbf{r}}
=\beta \sum_{\alpha} e^{i(\textbf{K}+\textbf{G})\cdot\textbf{r}} \bigg[\tilde{w}(\textbf{R}_{\alpha}) - \sum_{\textbf{G}'} W(\textbf{G}') e^{-i (\textbf{K}+\textbf{G}')\cdot\textbf{R}_{\alpha}} \bigg]\delta(\textbf{r}-\textbf{R}_{\alpha}). 
\end{eqnarray}

\noindent Taking the area integral over the entire unit cell of area $A_c$ leads to
\begin{eqnarray}
\sum_{\textbf{G}'} \bigg[D \abs{\textbf{K}+\textbf{G}'}^4&-&\omega^2 \rho h \bigg] W(\textbf{G}') \iint\limits_{A_c} e^{-i(\textbf{G}'-\textbf{G})\cdot\textbf{r}}dr^2 \nonumber \\
&=& \beta \sum_{\alpha} \bigg[\tilde{w}(\textbf{R}_{\alpha}) - \sum_{\textbf{G}'} W(\textbf{G}') e^{-i (\textbf{K}+\textbf{G}')\cdot\textbf{R}_{\alpha}} \bigg] \iint\limits_{A_c} e^{i(\textbf{K}+\textbf{G})\cdot\textbf{r}} \delta(\textbf{r}-\textbf{R}_{\alpha}) dr^2.
\end{eqnarray}

\noindent We now use the following relations 
\begin{subequations}
\begin{eqnarray}
\iint\limits_{A_c} e^{-i(\textbf{G}'-\textbf{G})\cdot\textbf{r}}dr^2 &=& \begin{cases}
    A_c,& \text{if } \textbf{G}=\textbf{G}'\\
    0,              & \text{otherwise}
\end{cases} \\
 \iint\limits_{A_c} f(\textbf{r}) \delta(\textbf{r}-\textbf{R}_{\alpha}) dr^2 &=& f(\textbf{R}_{\alpha})
\end{eqnarray}
\end{subequations}

\noindent to obtain
\begin{eqnarray}
A_c \bigg[D \abs{\textbf{K}+\textbf{G}}^4&-&\omega^2 \rho h \bigg] W(\textbf{G})  = \beta \sum_{\alpha} \bigg[\tilde{w}(\textbf{R}_{\alpha}) - \sum_{\textbf{G}'} W(\textbf{G}') e^{-i (\textbf{K}+\textbf{G}')\cdot\textbf{R}_{\alpha}} \bigg] e^{i(\textbf{K}+\textbf{G})\cdot\textbf{R}_{\alpha}}. 
\end{eqnarray}

Using the Bloch's theorem for the resonators, we write $\tilde{w}(\textbf{R}_{\alpha}) =\tilde{w}(\textbf{0}_{\alpha}) e^{-i \textbf{K}\cdot\textbf{R}_{\alpha}} $, where $\tilde{w}(\textbf{0}_{\alpha}) $ represents the Bloch displacement of the resonator (indexed with $\alpha$) at the reference unit cell. Thus, we deduce 
\begin{eqnarray}
\bigg[ a^4 \abs{\textbf{K}+\textbf{G}}^4 -  \mathrm{\Omega}^2 \bigg ] W(\textbf{G}) =  \gamma \mathrm{\Omega}_r^2  \sum_{\alpha}  e^{i \textbf{G}\cdot\textbf{R}_{\alpha}} \bigg[\tilde{w}(\textbf{0}_{\alpha})  - \sum_{\textbf{G}'} W(\textbf{G}') e^{-i \textbf{G}'\cdot\textbf{R}_{\alpha}}  \bigg]. \label{Eig1}
\end{eqnarray}

\noindent Similarly, we simplify the second governing Eq.~\eqref{EOM2} (for $\alpha=1,2,...,6$) as
\begin{eqnarray}
-\mathrm{\Omega}^2 \tilde{w}(\textbf{0}_{\alpha}) = \mathrm{\Omega}_r^2 \bigg[ \sum_{\textbf{G}} W(\textbf{G}) e^{-i \textbf{G}\cdot\textbf{R}_{\alpha}}-\tilde{w}(\textbf{0}_{\alpha}) \bigg].  \label{Eig2}
\end{eqnarray}

Given the $N \times N$ size of the reciprocal space, we arrange Eq.~\eqref{Eig1} and Eq.~\eqref{Eig2} in the form of an eigenvalue problem to solve for $\mathrm{\Omega}$ at a specific Bloch wave vector $\textbf{K}$ and obtain the dispersion relation. Note that we multiply Eq.~\eqref{Eig2} with $\gamma$ to make the matrices Hermitian. Therefore, we have
\begin{equation}
\begin{bmatrix} \textbf{P}_{11} & \textbf{P}_{12} \\ \textbf{P}_{21} & \textbf{P}_{22} \end{bmatrix} \left\{ \begin{array}{c} W(\textbf{G}) \\ \tilde{w}(\textbf{0}_{\alpha}) \end{array} \right \} = \mathrm{\Omega}^2 \begin{bmatrix} \textbf{Q}_{11} & \textbf{Q}_{12} \\ \textbf{Q}_{21} & \textbf{Q}_{22} \end{bmatrix} \left\{ \begin{array}{c} W(\textbf{G}) \\ \tilde{w}(\textbf{0}_{\alpha}) \end{array} \right \}
\end{equation}
\noindent with
\begin{eqnarray}
 \textbf{P}_{11}&=&a^4 \begin{bmatrix} \abs{\textbf{K}+\textbf{G}_1}^4 & 0 & \cdots & 0 \\ 0 & \abs{\textbf{K}+\textbf{G}_2}^4 & \cdots & 0\\  \vdots & \vdots & \ddots & \vdots \\ 0 & \cdots & 0 & \abs{\textbf{K}+\textbf{G}_{N^2}}^4\end{bmatrix} \nonumber \\ 
& +& \gamma \mathrm{\Omega}_r^2 \exp \left\{ i \begin{bmatrix} \textbf{G}_1 \\ \textbf{G}_2 \\ \vdots \\ \textbf{G}_{N^2} \end{bmatrix}  \begin{bmatrix} \textbf{R}_1 & \textbf{R}_2 &  \cdots & \textbf{R}_{6} \end{bmatrix} \right\} \exp \left\{-i \begin{bmatrix} \textbf{R}_1 \\ \textbf{R}_2 \\  \vdots \\ \textbf{R}_{6} \end{bmatrix}  \begin{bmatrix} \textbf{G}_1 & \textbf{G}_2 &  \cdots & \textbf{G}_{N^2} \end{bmatrix} \right\} \nonumber, \\
 \textbf{P}_{12}&=&\textbf{P}_{21}^{\dagger}=-\gamma \mathrm{\Omega}_r^2 \exp \left\{ i \begin{bmatrix} \textbf{G}_1 \\ \textbf{G}_2 \\ \vdots \\ \textbf{G}_{N^2} \end{bmatrix}  \begin{bmatrix} \textbf{R}_1 & \textbf{R}_2 &  \cdots & \textbf{R}_{6} \end{bmatrix} \right\}, \ \textbf{P}_{22} = \gamma \mathrm{\Omega}_r^2 \textbf{I}_6, \nonumber \\
\textbf{Q}_{11} &=& \textbf{I}_{N^2}, \ \textbf{Q}_{12} = \textbf{Q}_{21}^{\dagger} = \textbf{0}_{\{N^2,6\}}, \ \textbf{Q}_{22} = \gamma \textbf{I}_6, \nonumber
\end{eqnarray}

\noindent where `$\exp$', $\dagger$, $\textbf{I}$, and $\textbf{0}$ represent the exponential function, conjugate transformation, the identity matrix, and the null matrix, respectively. We choose $N=7$ for further calculations.

\begin{figure}[t]
\centering
\includegraphics[width=5in]{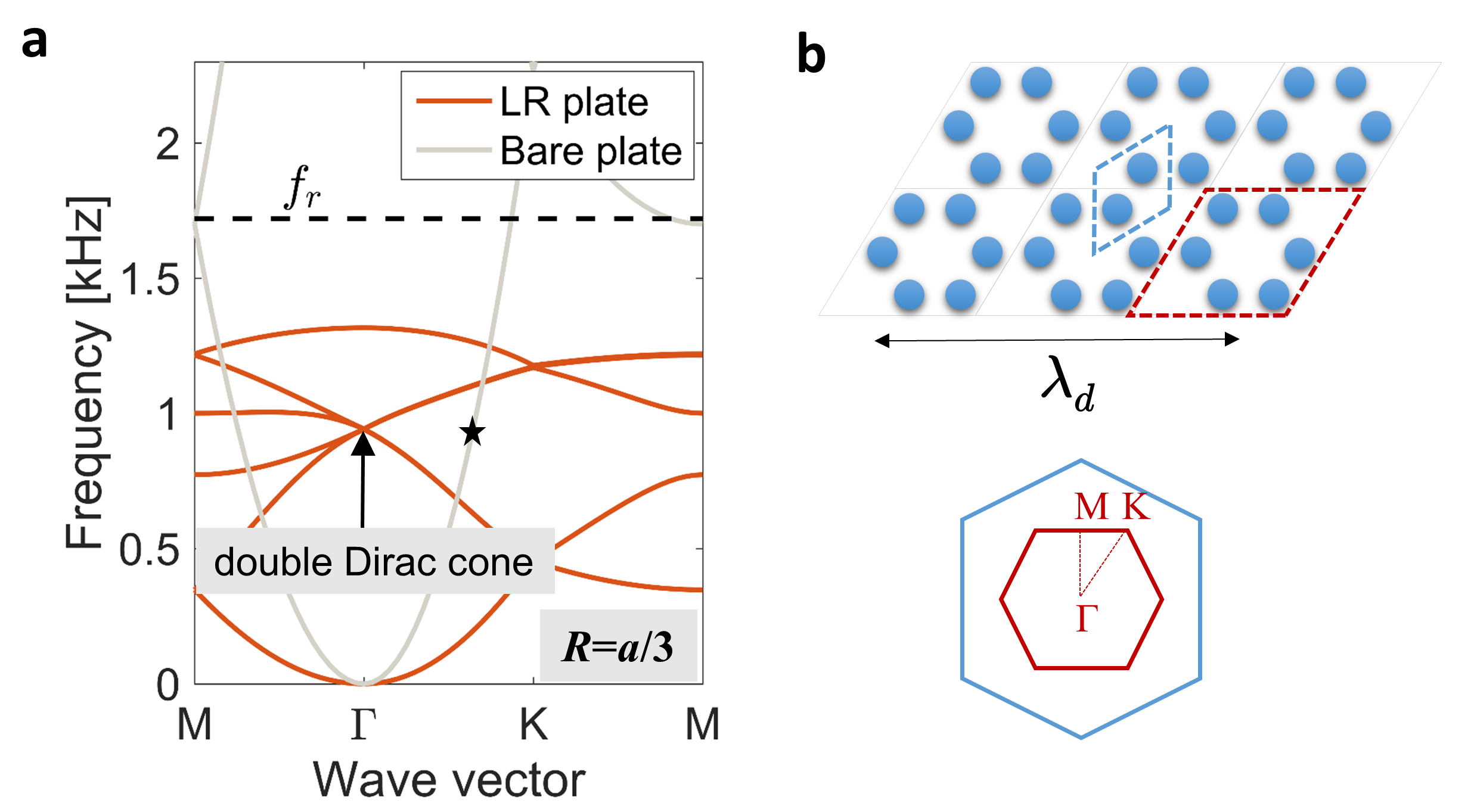}
\caption{(a) Dispersion plot for the LR plate with a hexagonal arrangement, i.e., $R=a/3$ (red curves). It shows a double Dirac cone, which is lower than the resonating frequency $f_r$ of the local resonators. Dispersion of the bare plate with no local resonators is in light gray color for comparison. (b) Two unit cell representations to enable the zone-folding (i.e., mathematical folding) of bands and the corresponding Brillouin zones below. The smaller cell (enclosed by blue dashed lines) represents the typical, irreducible unit-cell configuration for the hexagonal arrangement, while in this study, we consider the bigger unit cell (enclosed by red dashed lines) to create a double Dirac cone and for further topological manipulations. Their sizes are compared with the wavelength in the bare plate at the Dirac frequency (star mark)---making it a subwavelength unit design.}
\label{fig2}
\end{figure} 

\section{Band folding and subwavelength unit design}
By using the aforementioned technique, we calculate the dispersion relation for the hexagonal arrangement of resonators, i.e., $R=a/3$ (see Fig.~\ref{fig1}a) and plot in Fig.~\ref{fig2}a. Torrent \textit{et al.} \citep{Torrent2013} showed the existence of a single Dirac cone in such a system. Building on this finding, in the current study, we create a \textit{double} Dirac cone (two Dirac cone dispersion curves superimposed) with the frequency $f_d = 0.94$ kHz at the $\mathrm{\Gamma}$ point. This is possible because we have chosen a bigger unit cell consisting of six resonators instead of two (compare the unit cells in Fig.~\ref{fig2}b of different colors and corresponding Brillouin zones below). This results in the dispersion curves folded onto a smaller Brillouin zone \citep{Wu2015}. Note that the physics is the same in both representations and it is simply a \textit{mathematical} zone-folding of bands. However, achieving a double Dirac cone---which is a key ingredient of the spin Hall systems---guides us to realize topological effects by manipulating the geometrical configuration of the larger unit cell (to be further discussed in the Section V). 

In Fig.~\ref{fig2}a, we also mark the resonating frequency $f_r$ of the local resonator. It equals $f_r=(1/2\pi) \sqrt{\beta/m}=1.72$ kHz. It is important to realize that $f_d \leq f_r$, as thoroughly investigated by Torrent \textit{et al.} \citep{Torrent2013}. Therefore, the resonator design can be used as a tuning knob to push the Dirac frequency further down in the dispersion relation. In the same figure, we also plot the dispersion relation for a bare plate (i.e., the identical plate as the substrate described in Section II, but without local resonators attached). 
This is to compare the wavelength ($\lambda_d$) of flexural wave in the bare plate if excited at the Dirac frequency. This is indicated by the star marker on the dispersion curve. For the chosen set of design parameters, $\lambda_d$ is approximately 2.3 times longer than the length of the large unit cell (i.e., $a$), and 4 times longer than the size of the small, irreducible unit cell ($a/\sqrt{3}$). Figure~\ref{fig2}b shows the relative sizes of the unit cells compared to this wavelength, indicating subwavelength units of the LR plate. Therefore, as the topological effects will be seen around the Dirac frequency, this opens up new pathways to controlling large-wavelengths flexural waves by using a relatively small substrate. Again, by further reducing the resonant frequency, it is possible to shift the Dirac point to even lower frequency regime, thereby making the plate design deep-subwavelength. However, practical challenges in designing such a system can limit the same, e.g., due to heavy resonating masses and soft neck structures.

\section{Band inversion and topology}
\begin{figure}[t]
\centering
\includegraphics[width=5in]{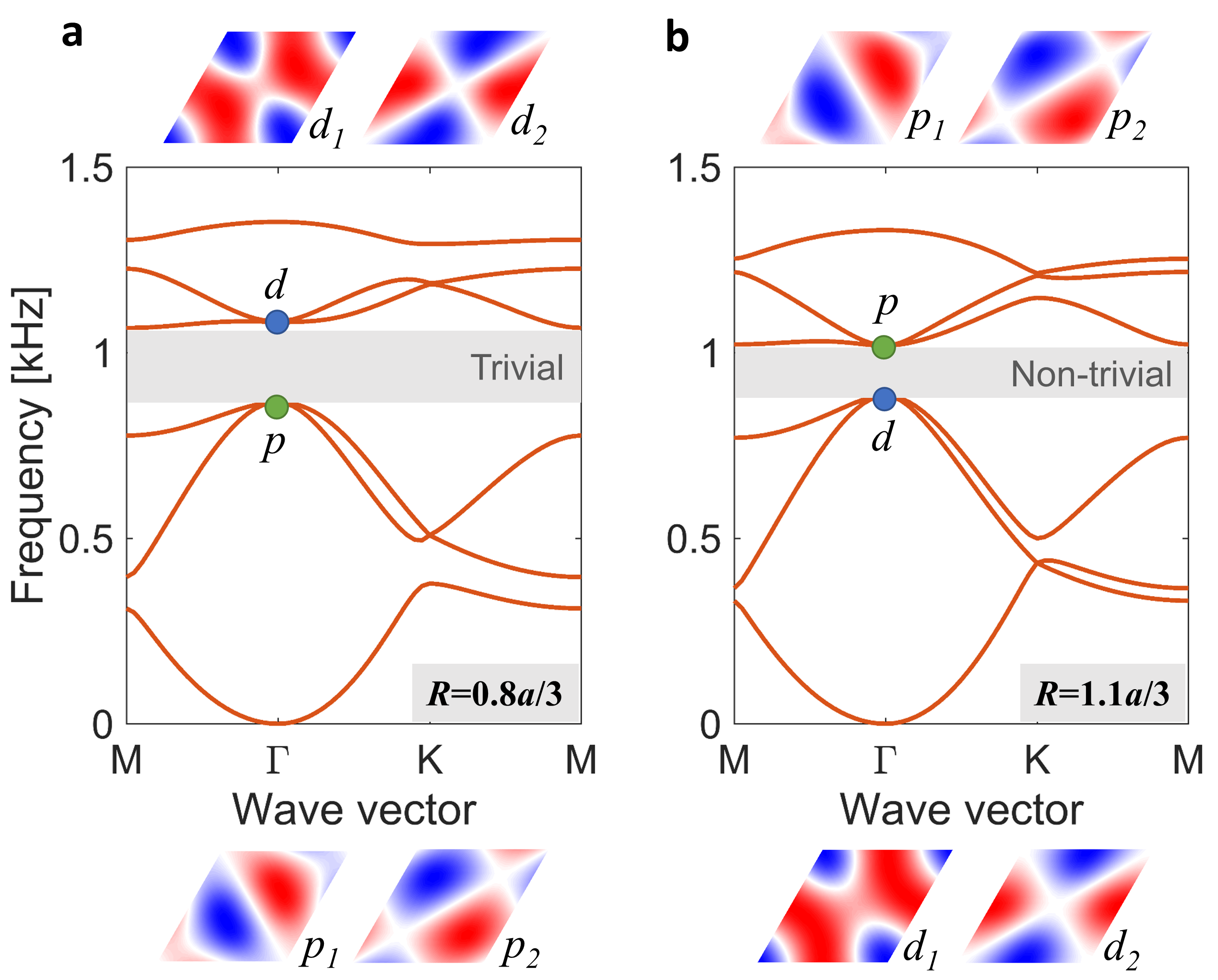}
\caption{Emergence of subwavelength Bragg band gaps and their inversion when the unit cell of the LR plate is perturbed around $R=a/3$ configuration. (a) A case with $R<a/3$ leading to topologically trivial band gap with $p$-type modes having lesser frequency than $d$-type modes. See the insets for the corresponding mode shapes with the colors indicating the out-of-plane displacements of the plate. (b) A case with $R>a/3$ leading to topologically non-trivial band gap with $p$- and $d$-type modes inverted.}
\label{fig3}
\end{figure} 

We now vary the radius $R$ and see its effects on the wave dispersion in the system. For $R<a/3$, as shown in Fig.~\ref{fig3}a, there emerges a band gap near the Dirac frequency. We call it a subwavelength Bragg band gap because it lies in the subwavelength regimes as discussed above but emerges due to the change in translational periodicity of the resonators. By keeping the $C_6$ symmetry intact, two modes on each side (lower or higher side) of the gap are degenerate at the $\mathrm{\Gamma}$ point. Seen in the insets are the corresponding degenerate mode shapes of the plate at the $\mathrm{\Gamma}$ point, which are obtained by the PWE method described in Section III. Here, the lower frequency modes are of $p$-type ($p_1$ and $p_2$ as shown in the bottom panel of Fig.~\ref{fig3}a), and the higher frequency modes are $d$-type ($d_1$ and $d_2$, upper panel in Fig.~\ref{fig3}a) as per the analogy to electronic orbital shapes. For $R>a/3$, however, the band gap still exists, but its topological characteristic is different from the earlier case. As shown in Fig.~\ref{fig3}b, the degenerate modes are flipped, i.e., $d$-type modes are at the lower frequency compared to $p$-type modes. This \textit{band inversion} as we vary $R$ around $R=a/3$  indicates a typical topological transition in the system. Again, the validity of this result based on the lumped mass model is verified and discussed in Appendix A in comparison with the FEM (using COMSOL Multiphysics), which takes into account all geometrical features in the resonator design.

The presence of degenerate modes around the band gap has important implications in realizing pseudo-spin Hall effect. One can take linear combinations of these modes and construct two alternate modes, i.e., pseudo-spin modes, without changing the physics of the system. Let $p_{\pm}=p_1\pm i p_2$ and $d_{\pm}=d_1\pm i d_2$ represent such spin modes for these degenerate points. The sign in the middle determines if these are rotating clockwise or counterclockwise. We can interpret the dispersion near the $\mathrm{\Gamma}$ point in terms of the pair of spins by projecting the eigenstates onto the spin basis \{$p_{\pm},d_{\pm}$\}. Therefore, the effective Hamiltonian of the system around the $\mathrm{\Gamma}$ point reduces to the one for Cd/Te/HgTe/CdTe quantum well \citep{Bernevig2006} and would resemble a mechanical pseudo-spin Hall system. One can show that the bands have non-zero spin Chern number for the case with $R>a/3$, hence, proving it to be topologically non-trivial \citep{Wu2015, Zhang2017, Yves2017}. 

\section{Emergence of topological interface state}
\begin{figure}[t]
\centering
\includegraphics[width=5in]{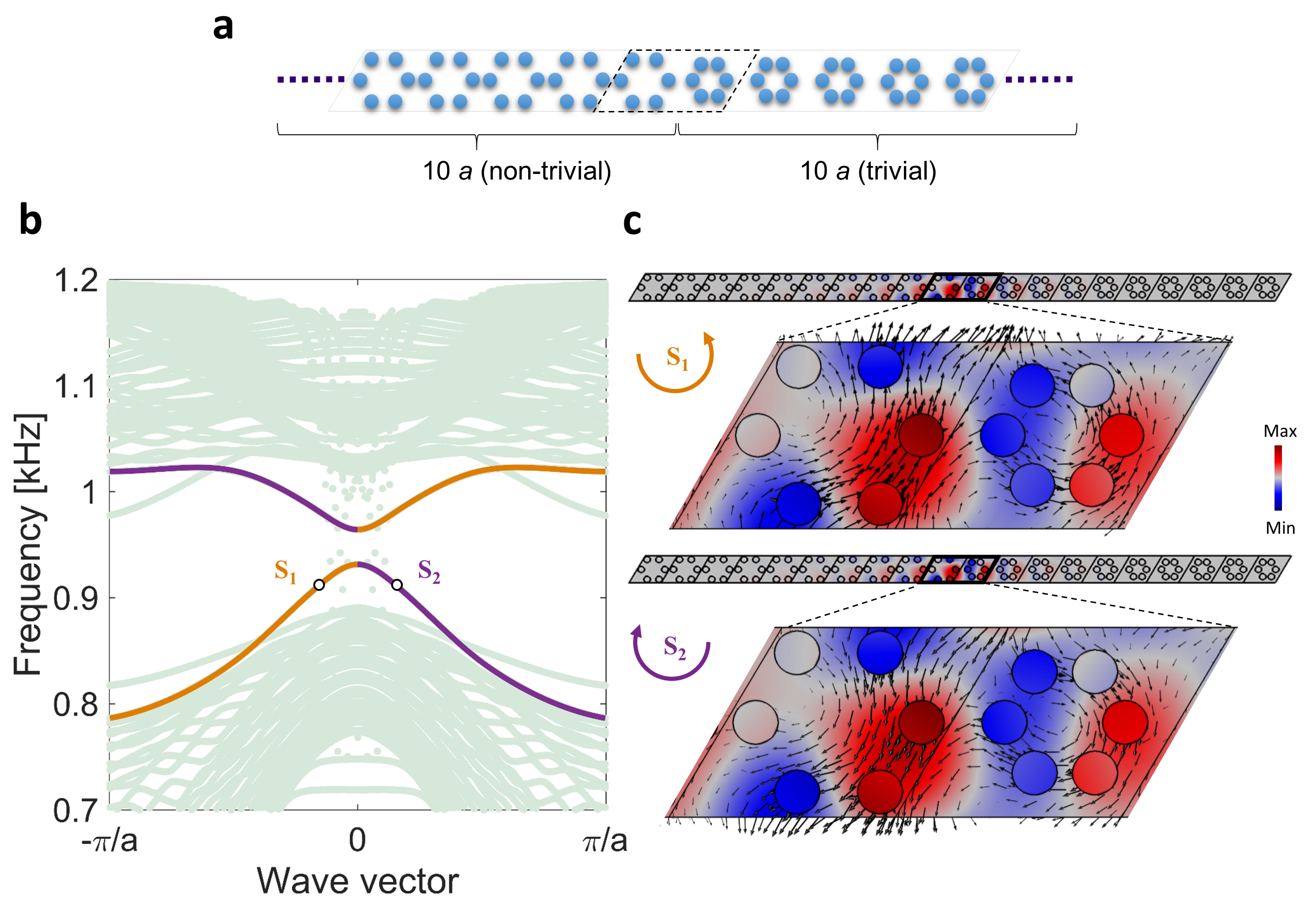}
\caption{(a) A supercell made by placing topologically distinct lattices adjacently. (b) Eigenfrequencies of the supercell as a function of the wave number in the periodic direction by analyzing a full-scale design with the FEM. We highlight clockwise (purple) and counterclockwise (yellow) pseudo-spin modes localized at the interface between the trivial and non-trivial lattices. In background are the other modes, i.e., bulk modes, and local modes at the extreme left and right end of the supercell. (c) Pseudo-spin mode shapes corresponding to the points S$_1$ and S$_2$ in (b). The color intensity represents the out-of-plane displacement, and the arrows indicate the time-averaged mechanical energy flux, thereby confirming their spin nature.}
\label{fig4}
\end{figure}

Now that we verified the feasibility of the double Dirac cone formation and the band inversion in the unit-cell level, we move to the investigation of wave guiding characteristics in multi-cell configurations. To account for more complicated geometry and boundary conditions in such a multi-cell setting, we resort to the FEM henceforth. According to the bulk-boundary correspondence of topology \citep{Hasan2010}, we expect distinct behaviors on the boundaries of topologically trivial and non-trivial lattices. One way to observe it clearly is to have topologically distinct lattices placed adjacently and investigate their connecting interface for a non-trivial local response. To this end, we take a supercell, which consists of both topologically trivial ($R=0.8a/3$) and non-trivial ($R=1.1a/3$) lattices, 10 units of each placed as one strip (Fig.~\ref{fig4}a). The periodic boundary condition is introduced in the direction of another lattice vector (at $60^{\circ}$ from the horizontal). In this way, such a system provides a quick way to calculate vibration responses at the interface and monitor their propagation along the periodic direction.

In Fig.~\ref{fig4}b, we plot the eigenfrequencies of the supercell as a function of wave number in the periodic direction. The presence of two modes inside the band gap (shown in purple and yellow colors) is especially striking, since those have the following non-trivial properties.  First, they represent two types of pseudo-spin modes localized at the interface: one rotates clockwise, while the other rotates counterclockwise. Second, both have opposite group velocities at a given frequency. Figure~\ref{fig4}c shows the respective mode shapes corresponding to points S$_1$ and S$_2$ in Fig.~\ref{fig4}b, which are excited at 0.91 kHz. Opposite spins and group velocities of these modes can be verified by looking at the harmonic evolution of these modes (see Supplementary Movie 1). We also plot in-plane time-averaged mechanical energy flux ($I_j=-\sigma_{ij}v_j$, where $\sigma_{ij}$ and $v_j$ are stress tensor and velocity vector, respectively) over a harmonic cycle as black arrows. This further confirms the spin nature of these flexural modes in the LR plate. 
There is a small frequency gap at the $\mathrm{\Gamma}$ point for these spin modes. The absence of topological interface modes indicates the absence of topological protection, and that suggests these pseudo-spin modes are not topologically protected in the full frequency band gap. This is because, in our system, the protection is guaranteed by the $C_6$ symmetry, which we break by introducing a sharp interface between topologically trivial and non-trivial lattices, and thereby resulting in an avoided crossing at the $\mathrm{\Gamma}$ point. Nevertheless, we will show in the next section that these modes can still be used to build robust and directional waveguides. The remedy to reduce the gap at the $\mathrm{\Gamma}$ point is to minimize the effect of the $C_6$ symmetry breaking at the interface. This can be done in several ways, including (1) by choosing the radii of trivial and non-trivial configurations as close as possible, or (2) by constructing a \textit{graded} interface between two topologically distinct lattices (see Appendix B). This therefore leads to a `greater' degree of protection of the topological spin-modes.

\section{Directional waveguides}
\begin{figure}[t]
\centering
\includegraphics[width=6in]{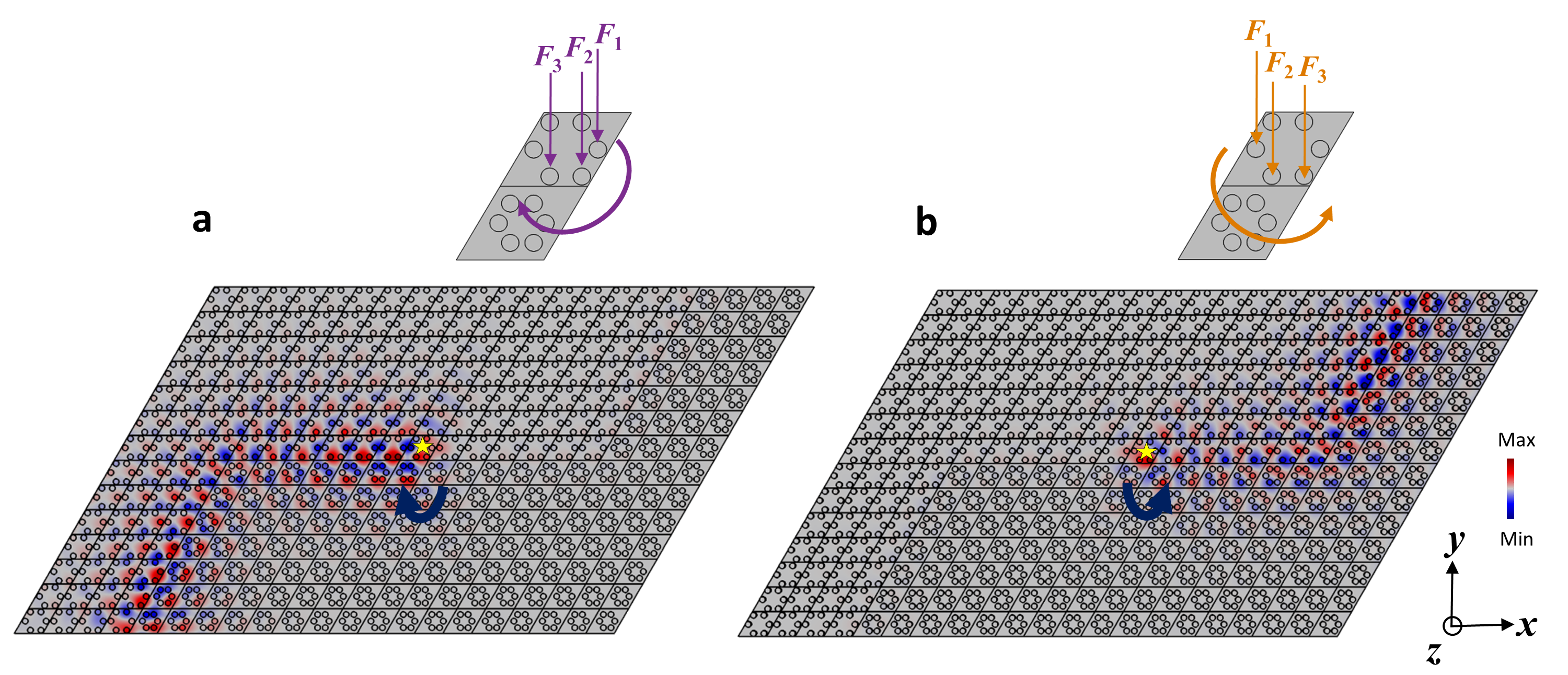}
\caption{Pseudo-spin dependent one-way propagation of flexural wave. The star indicates the zone of excitation. Multi-point phased excitation in force is used for exciting (a) clockwise and (b) counterclockwise spin modes. The color intensity represents the out-of-plane displacement. No back-scattering around the bends is observed.}
\label{fig5}
\end{figure} 

In the previous section, we have shown that the mechanical spin Hall effect enables us to have two pseudo-spins at one frequency but in the opposite directions. In order to demonstrate how this property can be used to build unconventional waveguides on plates, we combine topologically trivial and nontrivial LR plates to form a 2D structure. 
Figure~\ref{fig5} shows the waveguide (three linear segments with two bends) along the interface of two types of lattices. We give a forced excitation in the $z$-direction at the center of the plate (indicated by the star symbol) in such a way that we selectively excite spin modes. This could be done, for example, by choosing multiple points in the vicinity but with a phase difference in their forcing. Insets show the three excitation points in a nontrivial unit cell (i.e., with $R=1.1a/3$) where the spin is predominantly $d$-type. We extract the phase information from the spin modes in the supercell analysis done earlier for 0.91 kHz (see the resonators in red and blue colors, representing out-of-phase oscillations in Fig.~\ref{fig4}c). We apply this phased excitation as $F_1=F\exp(i\omega t)$, $F_2=F\exp(i\omega t+2\pi/3)$, and $F_3=F\exp(i\omega t+\pi)$. Note that the phase differences in the three excitation points are not equally spaced, but show $\pi/3$ and $2\pi/3$ differences between the neighboring ones (i.e., $\pi/3$ between $F_2$ and $F_3$, and $2\pi/3$ between $F_1$ and $F_2$). This excitation tactic induces clockwise spin in Fig.~\ref{fig5}a and counterclockwise spin in Fig.~\ref{fig5}b. We enforce low-reflecting boundary conditions on the plate and perform harmonic analysis using the FEM. 

We confirm the unique features of this topological waveguide. The clockwise spin mode propagates to the left (Fig.~\ref{fig5}a) and the counterclockwise spin propagates only to the right (Fig.~\ref{fig5}b). These spin waves propagate robustly along the waveguide interface in a way that even though there are sharp bends, there is no back-scattering and the spins remain intact (see Supplementary Movie 2). 
These simulation results imply that by using the pseudo-spin Hall effect induced in this LR plate structure, we can guide flexural waves in a selected path and direction without resorting to the breakage of time-reversal symmetry. That is, without using any active components, we can achieve directional control of low-frequency flexural waves simply by creating a topological boundary and exciting the host medium strategically via a phased excitation. It is important to note that robustness of these spin waves shown along a waveguide with sharp bends does not imply that these are also robust against any other types of `defects' along the waveguide as shown in acoustics recently \citep{Deng2017}. Though our elastic LR plate structure demands a thorough stand-alone study on this subject in future, we have explored exemplary cases here, in which certain defects along the waveguides can(not) back scatter these spin waves (Appendix C).
 
\section{Conclusions}
We have proposed a locally resonant plate structure to demonstrate the pseudo-spin hall effect for directional control of flexural waves. We show that the resonator design can be simplified with a lumped mass model and solved by employing the plane wave expansion method. This method enables us to efficiently investigate the key design parameters responsible for forming a double Dirac cone at a lower frequency than the resonating frequency of the local resonators. 
Keeping the $C_6$ symmetry intact, we perturb the unit cell and show an opening of subwavelength Bragg band gaps and the corresponding band-inversion process. This provides us with two topologically distinct lattice configurations. When these lattices are placed adjacently, we show the existence of two pseudo-spin modes traveling in the opposite directions along the interface. This unique feature is used to build topological waveguides with multiple bends and robustly guiding the spin-dependent flexural waves in a selected direction. 
The finding could be useful in designing compact and robust one-way channels for guiding low-frequency flexural waves in applications such as energy harvesting, sensing, and impact mitigation. Future studies include the optimization of the locally resonant plate configurations by using the proposed numerical techniques, as well as the experimental verification of the waveguiding effects, which will be reported by the authors' future publications.     

\begin{acknowledgments}
We gratefully acknowledge fruitful discussions with Krishanu Roychowdhury (Cornell University), Rui Zhu (Beijing Institute of Technology), Raj Kumar Pal (Georgia Institute of Technology), Simon Yves (CNRS), Romain Fleury (EPFL), Cheng He (Nanjing University), Zhiwang Zhang (Nanjing University), and Panayotis Kevrekidis (University of Massachusetts, Amherst). We are grateful for the support from NSF (CAREER-1553202 and EFRI-1741685). 
\end{acknowledgments}

\appendix
\section{COMPARISON BETWEEN THE LUMPED MASS MODEL (PWE) AND FULL-SCALE MODEL (FEM)}
\begin{figure}[h]
\centering
\includegraphics[width=5in]{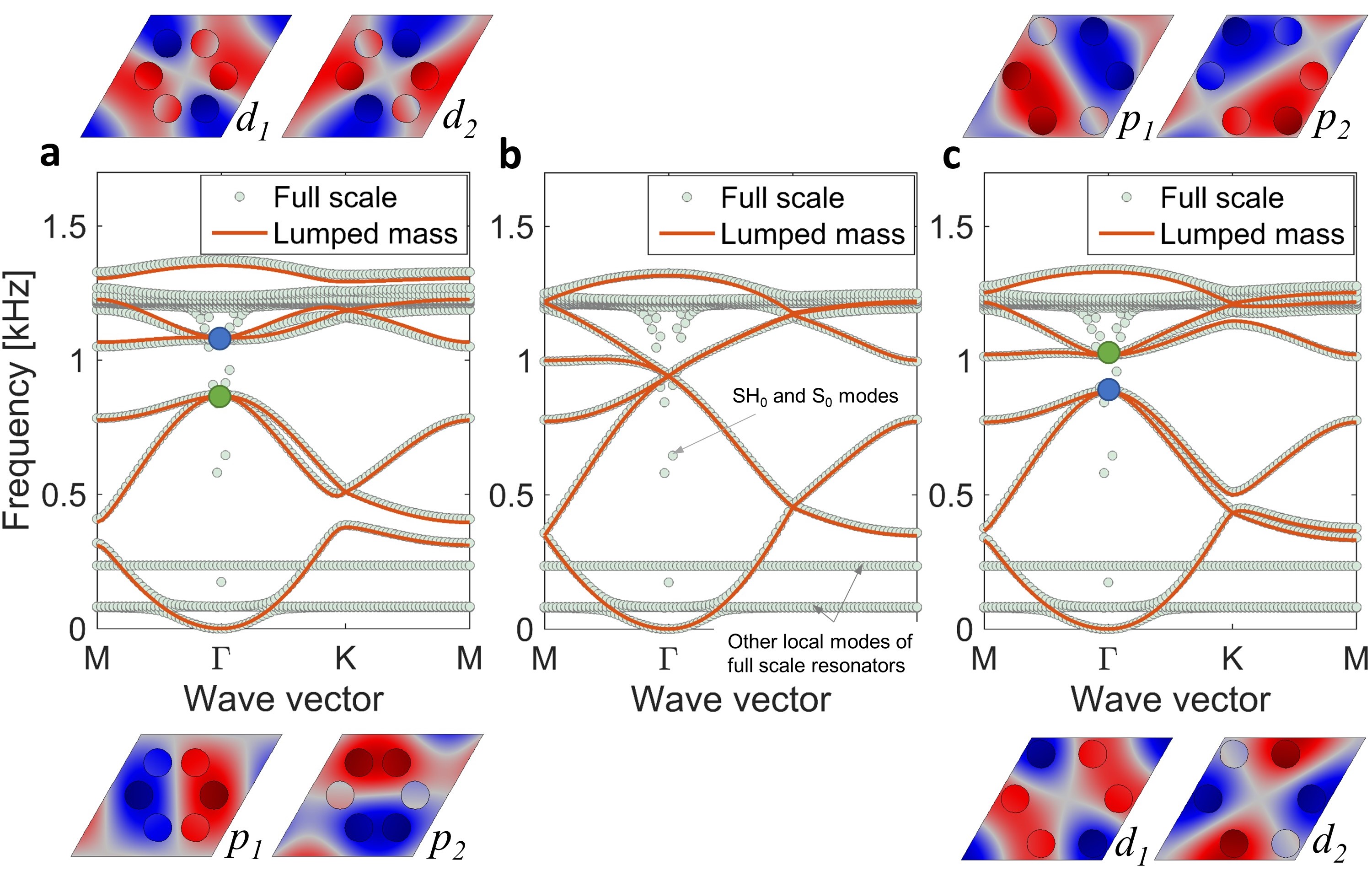}
\caption{Comparison of unit cell dispersion curves obtained from the lumped mass model (solid curves) and the full-scale FEM (dotted curves) for the cases: (a) $R=0.8a/3$, (b) $R=a/3$, and (c) $R=1.1a/3$. Excellent match is observed and band-inversion mechanism (insets with modes shapes obtained through full-scale FEM) is confirmed for flexural modes (A$_0$). Additional modes emerging in the full-scale numerical simulation are SH$_0$ and S$_0$ guided plate modes, and other modes of local resonators.}
\label{figS1}
\end{figure}

We corroborate the results obtained earlier based on the lumped mass model now by using the FEM, in which we account for all geometrical features described in Fig.~\ref{fig1}b. As shown in Fig.~\ref{figS1}, we clearly see an excellent agreement of the PWE method (red curves) with the FEM (green circles). It should be also noted that the $p$- and $d$-type degenerate mode shapes obtained through the lumped mass model comply with those obtained by the full-scale model (compare the inset images between Fig.~\ref{fig3} and Fig.~\ref{figS1}). In the dispersion relation obtained by the FEM, however, we observe that previously neglected shear-horizontal (SH$_0$) and shear (S$_0$) plate modes do appear for a thin plate. In our analysis, it is reasonable to ignore such modes and focus solely on anti-symmetric (A$_0$) flexural modes for the transverse source excitation at low frequencies as it was already demonstrated by full-scale simulations in Section VI and Section VII. We also see nearly flat dispersion curves due to the other modes of local resonators, which were not accounted for in the lumped mass model. Nevertheless, these modes have minimal coupling with the out-of-plane vibration of the plate, and these are away from the double Dirac cone. Therefore, the band-inversion process is not affected by them and it is reasonable to neglect them in the PWE method. 
 
 \section{GRADED INTERFACE BETWEEN TWO TOPOLOGICALLY DISTINCT LATTICES}
\begin{figure}[!]
\centering
\includegraphics[width=5in]{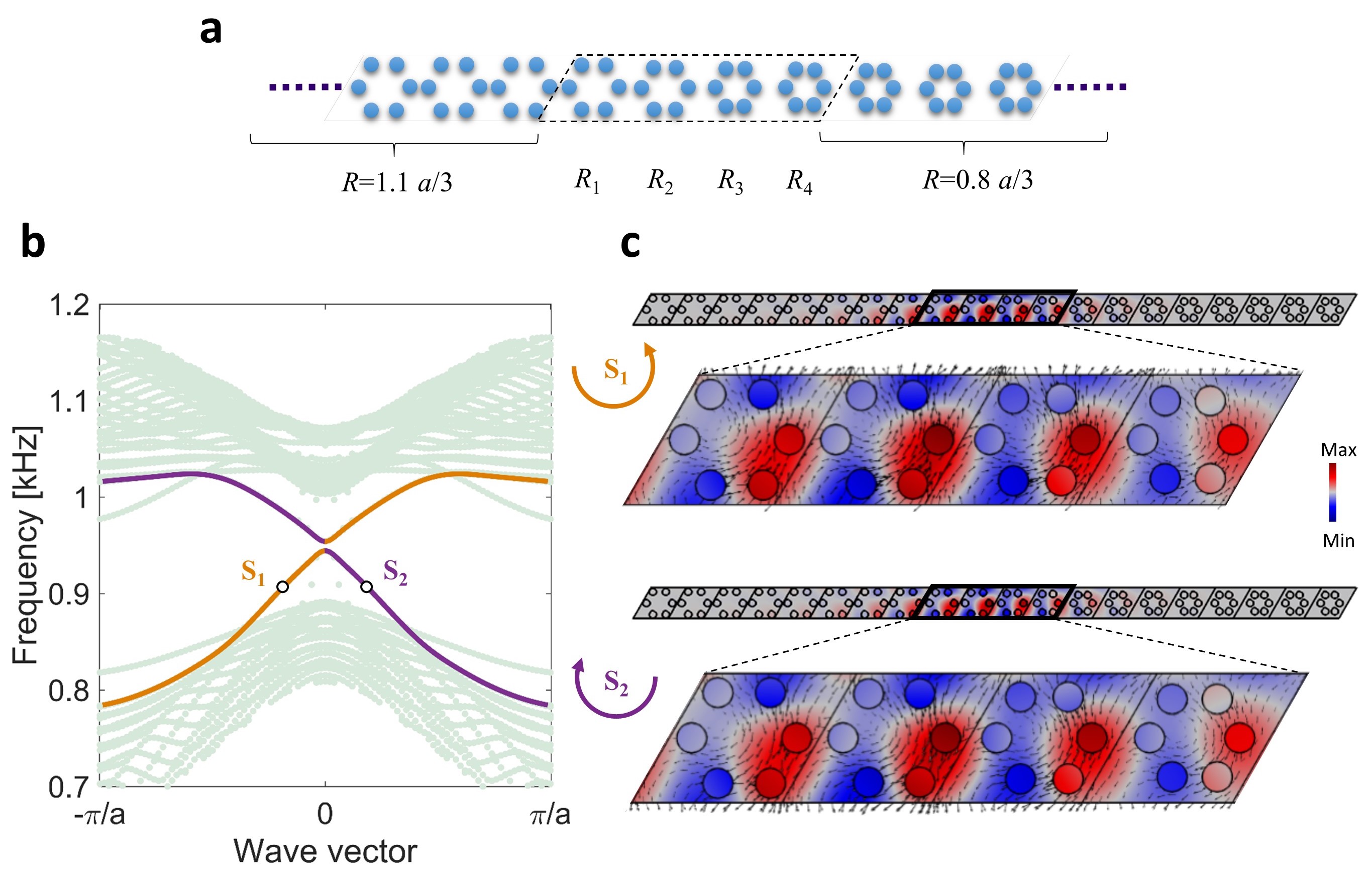}
\caption{The effect of a graded interface on the pseudo-spin modes at the interface. (a) A supercell with a smooth interface (radii $R_1>R_2>R_3>R_4$) between topologically trivial and non-trivial lattices. (b) Eigenfrequencies of the modified supercell. The gap reduction at the $\mathrm{\Gamma}$ point can be observed---indicating a greater degree of topological protection. (c) Pseudo-spin modes corresponding to S$_1$ and S$_2$ points in the dispersion curve.}
\label{figS2}
\end{figure}

Here we verify the scheme of reducing the gap observed at the $\mathrm{\Gamma}$ point in Fig.~\ref{fig4}b. The gap emerges due to the breakage of the $C_6$ symmetry at the interface. Therefore, we minimize the effect of symmetry breakage by modifying the interface. Figure~\ref{figS2}a shows a supercell, in which a topologically non-trivial lattice ($R=1.1a/3$) smoothly transitions to a topologically trivial lattice ($R=0.8a/3$) via four lattices with a gradient in their radii, i.e., $R_1=1.05a/3$, $R_2=1.02a/3$, $R_3=0.95a/3$, and $R_4=0.9a/3$. The gap reduction is confirmed in Fig.~\ref{figS2}b. It is about 3 times lesser than the one observed in Fig.~\ref{fig4}b. Spin nature of these interface modes is also intact as shown in Fig.~\ref{figS2}c.  

\section{THE PRESENCE OF DEFECTS ALONG THE WAVEGUIDE}
\begin{figure}[h]
\centering
\includegraphics[width=6in]{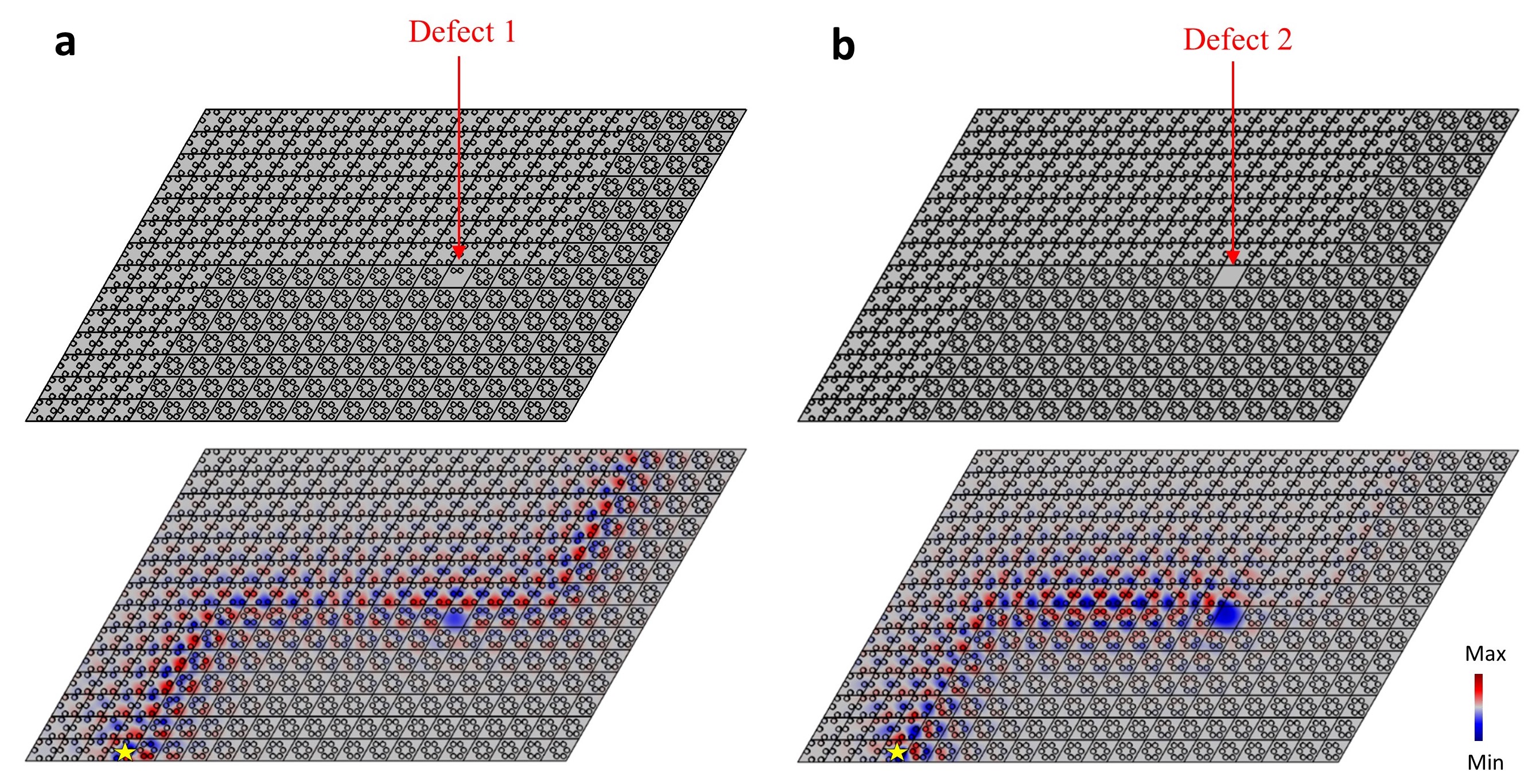}
\caption{The effect of defects on the spin-dependent one-way propagation of flexural waves. (a) Defect 1 is created by removing \textit{four} resonators in one of the unit cells along the interface. The star indicates the point of excitation. No obvious back-scattering is observed. (b) Defect 2 is created by removing \textit{all six} resonators in the same unit cell. We observe back-scattering of spin mode from this defect.}
\label{figS3}
\end{figure}

We further examine the robustness of spin-dependent one-way flexural waves against certain defects present at the topological interface as shown in Fig.~\ref{figS3}.
The first defect, i.e., Defect 1, is introduced by removing four local resonators from one of topologically trivial unit cells along the interface (see Fig.~\ref{figS3}a). We then excite a point on the left most end of the waveguide (shown by star marker) with low-reflecting boundary conditions and perform harmonic analysis for the same frequency as that in Fig.~\ref{fig5}. We observe that the presence of this defect does not have any noticeable effect in terms of scattering the spin mode. The mode continues to propagate from the left to the right with a counterclockwise spin (the same as in Fig.~\ref{fig5}b) and maintains similar modal amplitude before and after the defect location. Therefore, this defect would qualify as a non spin-mixing defect \citep{Deng2017}.   

\parindent=1.0em
We take another defect, i.e., Defect 2, in which we remove all six local resonators from the same unit cell (see Fig.~\ref{figS3}b). Thus this defect is `stronger' than Defect 1. Under the same excitation and boundary conditions as those in the previous case, we observe that this defect affects the spin mode more drastically. In Fig.~\ref{figS3}b, we see that the modal amplitude is almost negligible on the right side of the defect because of strong back-scattering. The defect causes the spin modes to mix. Consequently, the rightward-propagating counterclockwise spin mode is converted to leftward-propagating clockwise spin mode. This therefore represents a case of spin-mixing defect \citep{Deng2017}. 

\end{document}